# Conceptual Foundations of Diffusion in Magnetic Resonance

Cheng Guan Koay[1] and Evren Özarslan[2]

[1]Department of Medical Physics
University of Wisconsin School of Medicine and Public Health
Madison, WI 53705

[2]Department of Radiology
Brigham & Women's Hospital
Harvard Medical School
Boston, MA 02215

*Corresponding author:*
*Cheng Guan Koay, PhD*
*Department of Medical Physics*
*University of Wisconsin School of Medicine and Public Health*
*1161 Wisconsin Institutes for Medical Research (WIMR)*
*1111 Highland Avenue*
*Madison, WI 53705*
*E-mail: cgkoay@wisc.edu*

ABSTRACT

A thorough review of the q-space technique is presented starting from a discussion of Fick's laws. The work presented here is primarily conceptual, theoretical and hopefully pedagogical. We offered the notion of molecular concentration to unify Fick's laws and diffusion MRI within a coherent conceptual framework. The fundamental relationship between diffusion MRI and the Fick's laws are carefully established. The conceptual and theoretical basis of the q-space technique is investigated from first principles.

**INTRODUCTION**

Within the field of diffusion MRI, there are currently two widely accepted methods for imaging and probing tissue microstructure—diffusion tensor imaging (DTI) [1-2] and q-space imaging [3-6]. It is interesting to note that the essence of both of these techniques was already apparent at least in the context of NMR in the seminal paper[7] by Stejskal in 1965. The first detailed investigation on the relationships between diffusion tensor imaging and three-dimensional q-space imaging was performed by Basser [8].

The goal of this work is twofold. First, we will touch on the diffusion equation in various special cases leading up to the three-dimensional anisotropic diffusion equation and of the q-space technique, and employ the notion of concentration of diffusible molecules to highlight the important role played by Fick's laws in diffusion MRI. We will also present a detailed analysis of the relationship between DTI and q-space imaging. Finally, we will discuss recent developments related to q-space imaging.

## Section 1. From generalized Fick's law to three-dimensional anisotropic diffusion equation

We will begin with the generalized Fick's law and derive the three-dimensional anisotropic diffusion equation. In the next section, we will solve the three-dimensional anisotropic diffusion equation and point out the link to the propagator representation suggested by Kärger and Heink[9].

The generalized Fick's law describes how the molecular flux density, denoted by $\mathbf{j}(\mathbf{r},t)$, which is a vector quantity representing the number of molecules per unit oriented surface area normal to $\mathbf{j}(\mathbf{r},t)$ and per unit time, depends on the molecular concentration gradient, denoted by $\nabla C(\mathbf{r},t)$, and the geometry of the tissue or material microstructure influenced by the diffusion tensor, $\mathbf{D}$. Note that $C(\mathbf{r},t)$ is the number of molecules per unit volume; the components of $\nabla C(\mathbf{r},t)$ are the number of molecules per unit volume per unit length; the components of $\mathbf{D}$ have the dimensions of the squares of the length per unit time. Note that while the squares of the length has the same unit as the area, we should emphasize that conceptually the diffusion coefficient has nothing to do with an areal measurement or area but it is related to the second moment of the displacement probability; see Einstein's approach to diffusion equation [10-11]. Further, $\mathbf{r}$ is the position vector and $t$ denotes time.

The generalized Fick's law states that there is a flux of diffusible molecules from regions of high concentration to regions of low concentration but the flux is influenced by the geometry of the tissue or material microstructure as determined by the diffusion tensor, $\mathbf{D}$, and it is given by:

$$\mathbf{j}(\mathbf{r},t) = -\mathbf{D}\nabla C(\mathbf{r},t). \tag{1}$$

The diffusion tensor, $\mathbf{D}$, may have spatial and temporal dependence but for the sake of simplicity we will assume the diffusion tensor has no such dependence, which is equivalent to assuming that the possibly anisotropic microstructure of the tissue or material microstructure is homogeneous.

The differential fraction of the total number of molecules within the volume of interest, see Figure 1, can be obtained from the molecular concentration and is given by:

$$\frac{dN(t)}{M} = \frac{C(\mathbf{r},t)d^3r}{M},$$ (2)

where $d^3r$ is the volume element and $M \equiv \int_{whole\ space} C(\mathbf{r},t)d^3r$ is the total number of molecules in the whole space and is independent of time. Upon integration over the volume of interest, i.e., a subspace of the whole space, we have

$$\frac{N(t)}{M} = \frac{\int_{volume} C(\mathbf{r},t)d^3r}{M}.$$ (3)

Note that $N(t) \to M$ as the volume integration covers the whole space.

If there is no creation (source) or annihilation (sink) of molecules, say due to chemical reactions, within the volume of interest, the total number of molecules may still fluctuate as a result of molecules leaving or entering through the oriented surface element, $d\mathbf{S}$. Such fluctuation in the total number of molecules is due to thermal agitation. The rate of change of the total number of molecules within the volume of interest is related the molecular flux density by the following expression:

$$\frac{1}{M}\frac{dN(t)}{dt} = -\frac{1}{M}\oint_{surface} \mathbf{j}(\mathbf{r},t) \cdot d\mathbf{S},$$ (4)

$$= -\frac{1}{M}\int_{volume} \nabla \cdot \mathbf{j}(\mathbf{r},t)d^3r.$$ (5)

Note that the divergence theorem was used to obtain Eq. [5] from Eq. [4]. By equating the temporal derivative of the integrand of Eq. [3] and the integrand of Eq. [5], we arrive at the well-known (sink-less and source-less) continuity equation:

$$\frac{1}{M}\frac{\partial C(\mathbf{r},t)}{\partial t} + \frac{1}{M}\nabla\cdot\mathbf{j}(\mathbf{r},t) = 0 \qquad (6)$$

If creation of molecules is involved within the volume of interest, the right hand side of Eq. [6] will need to have a positive real function as a function of space and time; a negative real function is needed if annihilation is involved. Substituting Eq. [1] into Eq. [6], we arrive at the three-dimensional anisotropic diffusion equation [12]:

$$\frac{\partial}{\partial t}\left(\frac{C(\mathbf{r},t)}{M}\right) = \nabla\cdot\left(\mathbf{D}\nabla\left(\frac{C(\mathbf{r},t)}{M}\right)\right). \qquad (7)$$

We should note that Eq. [3] can be expressed in terms of the probability density function by defining $P(\mathbf{r},t) = C(\mathbf{r},t)/M$ so that $\int_{whole\ space} P(\mathbf{r},t)d^3r = 1$ and the evolution equation of $P(\mathbf{r},t)$ is described by the diffusion equation,

$$\frac{\partial P(\mathbf{r},t)}{\partial t} = \nabla\cdot\left(\mathbf{D}\nabla P(\mathbf{r},t)\right). \qquad (8)$$

In other words, instead of thinking $M$ molecules engage in the diffusion process, we think of the probability of finding a single molecule, which engages in a diffusion process, in a particular infinitesimal region of space at a particular time. Therefore, $P(\mathbf{r},t)$ and $C(\mathbf{r},t)$ are mathematically equivalent when we set $M = 1$.

## Section 2. The solution to the three-dimensional anisotropic diffusion equation and its conceptual link to the propagator representation

The derivation of the solution to the three-dimensional anisotropic diffusion equation shown in Eq. [8] is slightly more involved. All the essential steps needed to derive the solution of the three-dimensional anisotropic diffusion equation are given in three appendices. In Appendix A and Appendix B, we derive the solutions of the one-dimensional diffusion equation and the three-dimensional isotropic diffusion equation, respectively. In Appendix C, we derive the solution of the three-dimensional anisotropic diffusion equation by a means of a coordinate transformation to a special coordinate system in which diffusion is isotropic and the diffusion coefficient has the value of unity; the solution of this special diffusion equation, which was provided in Appendix B, is then transformed to the original coordinate system. The solution of Eq. [8] is shown to be:

$$P(\mathbf{r},t) = \frac{1}{(4\pi t)^{3/2}} \frac{1}{\sqrt{\det(\mathbf{D})}} \iiint P(\mathbf{r}',0) \exp\left(-\frac{(\mathbf{r}-\mathbf{r}')^T \mathbf{D}^{-1}(\mathbf{r}-\mathbf{r}')}{4t}\right) dx' dy' dz', \tag{9}$$

where $P(\mathbf{r}',0)$ is the initial condition for the probability density function $P(\mathbf{r},t)$ (equivalently, $C(\mathbf{r}',0)$ is the initial molecular concentration), and in the context of diffusion MRI, it is the probability density function of the tagged diffusible water molecules at the moment when the first magnetic field gradient pulse is applied. Further, when $t = \Delta$, which is the time interval between the two gradient pulses (see Figure 2), $P(\mathbf{r},t)$ should be thought of as the probability of finding the molecule at $\mathbf{r}$ after some time $t$ has elapsed with the initial probability density $P(\mathbf{r}',0)$.

If we assume that $P(\mathbf{r}',0) = \delta(\mathbf{r}'-\mathbf{r}_0)$, then we have

$$P(\mathbf{r},t) = \frac{1}{(4\pi t)^{3/2}} \frac{1}{\sqrt{\det(\mathbf{D})}} \exp\left(-\frac{(\mathbf{r}-\mathbf{r}_0)^T \mathbf{D}^{-1}(\mathbf{r}-\mathbf{r}_0)}{4t}\right). \tag{10}$$

The propagator, the term used by Kärger and Heink [9], which is a well known concept in quantum mechanics[13], is the multivariate Gaussian probability density function that appeared in the integrand of Eq. [9] and $P(\mathbf{r},t)$ is the average propagator because $P(\mathbf{r},t)$ is simply the expectation (or average) of the multivariate Gaussian probability density function with respect to the initial probability density function $P(\mathbf{r}',0)$. Another way of looking at the propagator that is more in line with the quantum mechanical approach[13] is to think of it as an operator that operates on $P(\mathbf{r}',0)$ to produce $P(\mathbf{r},t)$. In the context of a homogeneous system, it is interesting to note that as the initial probability density function, $P(\mathbf{r}',0)$, approaches $\delta(\mathbf{r}'-\mathbf{r}_0)$ the average propagator approaches the propagator.

If we assume that the system under investigation is very complex and that the assumption of spatial homogeneity may not be tenable in such a system, we should, therefore, replace the multivariate Gaussian probability density function in Eq. [9] with $P(\mathbf{r},t\,|\,\mathbf{r}',t=0)$:

$$P(\mathbf{r},t) = \iiint P(\mathbf{r}',0)P(\mathbf{r},t\,|\,\mathbf{r}',t=0)dx'dy'dz' \qquad (11)$$

Equivalently, if we set $\mathbf{R} = \mathbf{r}-\mathbf{r}'$, the average propagator may be written in terms of the dynamic displacement [5]:

$$P_D(\mathbf{R},t) \equiv \iiint P(\mathbf{r}',0)P(\mathbf{R}+\mathbf{r}',t\,|\,\mathbf{r}',t=0)dx'dy'dz'. \qquad (12)$$

$P_D(\mathbf{R},t)$ in Eq. [12] is the average propagator and $P(\mathbf{R}+\mathbf{r}',t\,|\,\mathbf{r}',t=0)$ is the propagator, which is the conditional probability density for finding a molecule at $\mathbf{R}+\mathbf{r}'$ after some time $t$, given the initial condition that it was at $\mathbf{r}'$ at $t=0$. Most importantly, we should point out that $P(\mathbf{r}',0)$ is the *a priori* probability density function of the tagged diffusible water molecules at the moment when the first magnetic field gradient pulse has just been applied, which is the topic we discuss next.

## Section 3. Complex representation of transverse magnetization, Stejskal-Tanner pulse sequence and normalized q-space signal representation

By definition, the magnetization vector, $\mathbf{M}(\mathbf{r},t) = [M_x(\mathbf{r},t), M_y(\mathbf{r},t), M_z(\mathbf{r},t)]^T$, is the vector sum of magnetic moments per unit volume, which is dimensionally the same as the molecular concentration. The complex representation of the transverse magnetization in a spin echo pulse sequence is usually expressed as follows:

$$M_+(\mathbf{r}(t),t) = M_+(\mathbf{r}_0,0)\exp(-\tfrac{t}{T_2})\exp(-i\omega_0 t)\exp(-i\int_0^t \omega(\mathbf{r}(\tau),\tau)d\tau)\,, \tag{13}$$

where $M_+(\mathbf{r},t) = M_x(\mathbf{r},t) + iM_y(\mathbf{r},t)$, $\omega_0$ is the Larmor angular frequency, $T_2$ is the spin-spin (or transverse) relaxation time. $\omega(\mathbf{r},\tau)$ may be thought of as the angular frequency due to several factors such as magnetic field inhomogeneity or the application of magnetic field gradient. In the context of q-space technique, $\omega(\mathbf{r},\tau)$ is due to the pulsed field gradient and is expressed as

$$\omega(\mathbf{r}(\tau),\tau) = \gamma\mathbf{G}(\tau)\cdot\mathbf{r}(\tau)\,. \tag{14}$$

Here, we have explicitly shown the time dependence in the position vector, $\mathbf{r}(t)$, in anticipation of later development. Note that the negative sign in the exponent that involves the angular frequency is consistent with the convention that if the projection of the gradient vector on the z-axis (the axis of the main magnetic field) is positive then the angular velocity associated with the gradient vector is pointing in the negative z-axis.

To facilitate the discussion of q-space technique, we show in Figure 2A the well-known Stejskal-Tanner pulse sequence. Here, we list the chronology of the behavior of the magnetization vector in a single cycle of the Stejskal-Tanner pulse sequence:

- At $t = t_1$, $M_+(\mathbf{r}(t_1), t_1) = M_+(\mathbf{r}_0, 0)\exp(-\frac{t_1}{T_2})\exp(-i\omega_0 t_1)$. Note that $M_+(\mathbf{r}_0, 0)$ is the *a priori molecular concentration* (or equivalently, upon normalization, the *a priori* probability density function) of the diffusible water molecules, i.e., $M_+(\mathbf{r}_0, 0) = C(\mathbf{r}_0, 0)\exp(i\theta)$ for some $\theta$.

- At $t = t_1 + \delta$, we have

$$M_+(\mathbf{r}(t_1 + \delta), t_1 + \delta)$$

$$= C(\mathbf{r}_0, 0)\exp(i\theta)\exp(-\frac{t_1 + \delta}{T_2})\exp(-i\omega_0(t_1 + \delta))\exp(-i\gamma\int_{t_1}^{t_1 + \delta}\mathbf{G}(\tau)\cdot\mathbf{r}(\tau)d\tau),$$

$$= C(\mathbf{r}_1, \tau = 0)\exp(i\theta)\exp(-\frac{t_1 + \delta}{T_2})\exp(-i\omega_0(t_1 + \delta))\exp(-i\gamma\delta\mathbf{G}\cdot\mathbf{r}_1), \qquad (15)$$

At $t = t_1 + \delta$, we assume that the concentration of tagged water molecules, $C(\mathbf{r}_1, \tau = 0)$, is the same as $C(\mathbf{r}_0, t = 0)$ because the initial concentration is uniform, i.e., there is no concentration gradient of (non-tagged) water molecules within the volume of interest prior to the gradient pulse, which created the tagged water molecules.

For convenience, we also used a different temporal variable, denoted $\tau$, in the initial molecular concentration and its subsequent evolution in time. We also assumed that the motion of the tagged molecules is negligible during the creation of the initial concentration of tagged molecules, which is during the application of the magnetic pulsed field gradient. This assumption is known as the narrow-pulse approximation. Therefore, the position vector is assumed to remain fixed, between $t_1$ and $(t_1 + \delta)$, and it is denoted by $\mathbf{r}_1$.

- From the first gradient pulse until $t = t_1 + \Delta + \delta$, we have to introduce the evolutionary aspect of the initial concentration of tagged water molecules into the magnetization equation together with the phase information affected by the refocusing RF pulse at TE/2, i.e., the expectation of $\exp(+i\gamma\delta G\mathbf{g}\cdot\mathbf{r}_1)$ with respect to the propagator, which may be expressed as

$$\left\langle \exp(+i\gamma\delta G\mathbf{g}\cdot\mathbf{r}_1)\right\rangle_C \equiv \iiint \exp(+i\gamma\delta G\mathbf{g}\cdot\mathbf{r}_1)C(\mathbf{r}_1,\tau=0)P(\mathbf{r}_2,\tau=\Delta\,|\,\mathbf{r}_1,\tau=0)dx_1dy_1dz_1 \ . \quad (16)$$

Note that the initial negative sign in $\exp(-i\gamma\delta G\mathbf{g}\cdot\mathbf{r}_1)$ has been changed to the positive because of the refocusing RF pulse.

Therefore, the complex transverse magnetization is given by:

$$M_+(\mathbf{r}(t_1+\Delta+\delta),t_1+\Delta+\delta)$$

$$=\Psi\left\langle \exp(+i\gamma\delta G\mathbf{g}\cdot\mathbf{r}_1)\right\rangle_C \exp\!\left(-i\gamma\int_{t_1+\Delta}^{t_1+\Delta+\delta}\mathbf{G}(\tau)\cdot\mathbf{r}(\tau)d\tau\right),$$

$$=\Psi\left\langle \exp(+i\gamma\delta G\mathbf{g}\cdot\mathbf{r}_1)\right\rangle_C \exp(-i\gamma\delta G\mathbf{g}\cdot\mathbf{r}_2)\,,$$

$$=\Psi\iiint \exp(+i\gamma\delta G\mathbf{g}\cdot\mathbf{r}_1)C(\mathbf{r}_1,\tau=0)P(\mathbf{r}_2,\tau=\Delta\,|\,\mathbf{r}_1,\tau=0)dx_1dy_1dz_1\exp(-i\gamma\delta G\mathbf{g}\cdot\mathbf{r}_2)\,,$$

$$=\Psi\iiint \exp(-i\gamma\delta G\mathbf{g}\cdot(\mathbf{r}_2-\mathbf{r}_1))C(\mathbf{r}_1,\tau=0)\,P(\mathbf{r}_2,\tau=\Delta\,|\,\mathbf{r}_1,\tau=0)\ dx_1dy_1dz_1 \ . \quad (17)$$

Note that $\Psi$ is a scaling factor that depends on $T_2$ decay, the exponential function involving the Larmor frequency and the arbitrary phase information in $\exp(i\theta)$. Similar to the previous section, it is assumed that the motion of the tagged molecule is negligible during the pulsation of the second pulsed field gradient. In other words, the position vector, $\mathbf{r}$, is a constant vector between $t_1+\Delta$ and $(t_1+\Delta+\delta)$, and is denoted by $\mathbf{r}_2$. Note that the refocusing pulse effectively flips the polarity of the first gradient. At $t=TE$, we have the echo formation. We note here that we are not interested in the motion of the magnetization between the second gradient pulse and the time of echo formation as such motion does not change the total magnetization. The only relevant information about the motion of the water molecules is in the interval between the two pulsed field gradients and it shows up as a variation in the phase. The q-space signal is obtained by first normalizing the magnetization by a division by $\Psi$ and then integrating over the volume of interest. In short, the q-space signal, denoted by $E_C(\mathbf{q}, \Delta)$, has following form:

$$E_C(\mathbf{q}, \Delta)$$

$$= \iiint \iiint C(\mathbf{r}_1, \tau = 0)\, P(\mathbf{r}_2, \tau = \Delta \mid \mathbf{r}_1, \tau = 0) \exp(-i\, 2\pi\, \mathbf{q} \cdot (\mathbf{r}_2 - \mathbf{r}_1))\; dx_1 dy_1 dz_1 dx_2 dy_2 dz_2 \;, \quad (18)$$

with $\mathbf{q} = \frac{1}{2\pi}\, \gamma \delta G \mathbf{g}$.

More compactly, we write

$$E_C(\mathbf{q}, \Delta) = \iint C(\mathbf{r}_1, \tau = 0)\, P(\mathbf{r}_2, \tau = \Delta \mid \mathbf{r}_1, \tau = 0) \exp(-i\, 2\pi\, \mathbf{q} \cdot (\mathbf{r}_2 - \mathbf{r}_1)) d\mathbf{r}_1 d\mathbf{r}_2 \qquad (19)$$

or

$$E(\mathbf{q}, \Delta) \equiv \frac{E_C(\mathbf{q}, \Delta)}{M} = \iint P(\mathbf{r}_1, \tau = 0)\, P(\mathbf{r}_2, \tau = \Delta \mid \mathbf{r}_1, \tau = 0) \exp(-i\, 2\pi\, \mathbf{q} \cdot (\mathbf{r}_2 - \mathbf{r}_1)) d\mathbf{r}_1 d\mathbf{r}_2 \qquad (20)$$

By a change of variables from $\mathbf{r}_1$, $\mathbf{r}_2$, to $\mathbf{r}$ and $\mathbf{R}$ given by

$$\mathbf{r} = \mathbf{r}_1 \text{ and } \mathbf{R} = \mathbf{r}_2 - \mathbf{r}_1,$$

and if we arrange the six dimensional variables in a form of a vector given by $[x, y, z, R_x, R_y, R_z] = [x_1, y_1, z_1, x_2 - x_1, y_2 - y_1, z_2 - z_1]$, it is easy to see that the six-dimensional Jacobian matrix is of the following form in block matrix notation:

$$\begin{pmatrix} \mathbf{I} & \mathbf{0} \\ -\mathbf{I} & \mathbf{I} \end{pmatrix},$$

where $\mathbf{0}$ is the 3 by 3 null matrix and $\mathbf{I}$ is the 3 by 3 identity matrix. It is also easy to see that the determinant of this Jacobian matrix is unity but care must be taken in dealing with the limits of integration of these variables to avoid having to subtract one positive infinity from another positive infinity; if we adopt the following strategy by integrating the components of $\mathbf{r}_1$ from positive infinity to negative infinity and of $\mathbf{r}_2$ from negative infinity to positive infinity, this strategy will produce a legitimate multi-dimensional integral and the negative signs due to reversal of limits of integration will eventually cancel among themselves on both sides of the integrations, one with $\mathbf{r}$ and $\mathbf{R}$ and the other with $\mathbf{r}_1$ and $\mathbf{r}_2$, so that both integrations are equivalent.

Hence,

$$E(\mathbf{q},\Delta) = \iint P(\mathbf{r},\tau=0)\,P(\mathbf{r}+\mathbf{R},\tau=\Delta\,|\,\mathbf{r},\tau=0)\,d\mathbf{r}\,\exp(-i\,2\pi\,\mathbf{q}\cdot\mathbf{R})\,d\mathbf{R}\,, \tag{21}$$

or

$$E(\mathbf{q},\Delta) = \int P_D(\mathbf{R},\Delta)\,\exp(-i\,2\pi\,\mathbf{q}\cdot\mathbf{R})\,d\mathbf{R}\,, \tag{22}$$

with

$$P_D(\mathbf{R},\Delta) \equiv \int P(\mathbf{r},0)P(\mathbf{R}+\mathbf{r},\tau=\Delta\,|\,\mathbf{r},\tau=0)d\mathbf{r}\,. \tag{23}$$

Equation [22] is the well known normalized q-space signal expression, which is related to the average propagator, $P_D(\mathbf{R},\Delta)$, via the forward Fourier transform relationship. Finally, $\mathbf{R}$ is known as the dynamic displacement [5]. Interested reader is urged to study Appendix D for further details.

**Section 4. Fick's laws in diffusion MRI**

In the context of diffusion MRI, it is very important to realize that the notion of molecular concentration is the concentration of *the tagged diffusible water molecules with the same phase* as shown in Figure 3. The almost instantaneous tagging of diffusible water molecules that lie on the three-dimensional plane with *the same phase* in effect introduces, for lack of a better name, a 'fictitious' concentration of diffusible water molecules, which may be thought of as a planar probability density function. Because of thermal agitation, there will be a concentration gradient of these tagged diffusible water molecules from the two-dimensional plane to the ambient space, which does not include diffusible water molecules with the same phase when the gradient strength is not very large. It is important to point out that the concentration of water molecules, tagged or non-tagged, may be relatively constant throughout the whole space.

**Section 5. The relationship between q-space and DTI**

Since the average propagator is a real-valued function, the real and imaginary parts of $E(\mathbf{q}, \Delta)$ have to be even and odd, respectively. In what follows, we will follow the analysis presented by Basser [8] and will point out the differences in our results from our those of Basser.

In the limit of large dynamic displacement, the asymptotic expression of the average propagator can be determined by the method of stationary phase. Thus, we are interested in the following expression:

$$P_D(\mathbf{R}, \Delta) \approx \int \exp(\ln(E_{real}(\mathbf{q}, \Delta))) \exp(+i 2\pi \mathbf{q} \cdot \mathbf{R}) d\mathbf{q} \qquad (\mathbf{R} \to \infty). \qquad (24)$$

Since the real part of $E(\mathbf{q}, \Delta)$, denoted by $E_{real}(\mathbf{q}, \Delta)$, is even and $E_{real}(\mathbf{0}, \Delta) = 1$, it is clear that the maximum of $E_{real}(\mathbf{q}, \Delta)$ is at $\mathbf{q} = \mathbf{0}$ and the major contribution to $P_D(\mathbf{R}, \Delta)$ would come from the Hessian matrix in the Taylor expansion of $E_{real}(\mathbf{q}, \Delta)$ about $\mathbf{q} = \mathbf{0}$. Therefore,

$$\ln(E_{real}(\delta\mathbf{q}, \Delta)) \approx \ln(E_{real}(\mathbf{0}, \Delta)) + \frac{1}{E_{real}(\mathbf{0}, \Delta)}(\delta\mathbf{q})^T . \nabla E_{real}(\mathbf{0}, \Delta)$$

$$+ \frac{1}{2E_{real}(\mathbf{0}, \Delta)}(\delta\mathbf{q})^T \nabla_H^2 E_{real}(\mathbf{0}, \Delta)\delta\mathbf{q}$$

$$- \frac{1}{2[E_{real}(\mathbf{0}, \Delta)]^2}(\delta\mathbf{q})^T \nabla E_{real}(\mathbf{0}, \Delta) . [\nabla E_{real}(\mathbf{0}, \Delta)]^T \delta\mathbf{q}$$

$$\approx \frac{1}{2}(\delta\mathbf{q})^T \nabla_H^2 E_{real}(\mathbf{0}, \Delta)\delta\mathbf{q} = -\frac{1}{2}(\delta\mathbf{q})^T \mathbf{G} \, \delta\mathbf{q}. \qquad (25)$$

In the equation above, the Hessian matrix of $f$ is denoted by $\nabla_H^2 f$. The negative Hessian matrix of the real part of the normalized q-space signal is denoted by $\mathbf{G}$. By substituting Eq. [25] into Eq. [24], we arrive at a multivariate Gaussian distribution:

$$P_D(\mathbf{R}, \Delta) \approx \int \exp(-\frac{1}{2}\mathbf{q}^T \mathbf{G} \, \mathbf{q}) \exp(+i 2\pi \mathbf{q} \cdot \mathbf{R}) d\mathbf{q}, \qquad (\mathbf{R} \to \infty)$$

$$= \frac{(2\pi)^{3/2}}{\sqrt{\det(\mathbf{G})}} \exp(-2\pi^2 \mathbf{R}^T \mathbf{G}^{-1} \mathbf{R}), \qquad (\mathbf{R} \to \infty). \qquad (26)$$

Since the average propagator used in diffusion tensor imaging is given by:

$$\frac{1}{(4\pi\Delta)^{3/2}\sqrt{\det(\mathbf{D})}}\exp(-\frac{\mathbf{R}^T\mathbf{D}^{-1}\mathbf{R}}{4\Delta}),$$ (27)

we need only to make the following the identification

$$\mathbf{G}=8\pi^2\mathbf{D}\Delta$$ (28)

in Eq. [26] to transform Eq. [26] in Eq. [27].

Note that Eq. [28] is different from that of Basser because he assumed $\mathbf{G}$ to be $2\mathbf{D}\Delta$, he interpreted $\mathbf{G}$ to be the mean-squared displacement, see Eq. [C7] and Eq. [C8] in Appendix C. By adopting the convention expressed in Eq. [28], it is clear that Eq. [26] reduces to the well-known multivariate Gaussian distribution used in diffusion tensor imaging:

$$P_D(\mathbf{R},\Delta)\approx\frac{1}{(4\pi\Delta)^{3/2}\sqrt{\det(\mathbf{D})}}\exp(-\frac{\mathbf{R}^T\mathbf{D}^{-1}\mathbf{R}}{4\Delta})\qquad(\mathbf{R}\rightarrow\infty).$$ (29)

From Eq. [25], it is easy to see that

$$E_{real}(\mathbf{q},\Delta)\approx\exp(-\frac{1}{2}\mathbf{q}^T\mathbf{G}\,\mathbf{q})=\exp(-4\pi^2\Delta\,\mathbf{q}^T\mathbf{D}\,\mathbf{q}).$$ (30)

**DISCUSSION**

We have to emphasize that the work presented here is primarily conceptual and theoretical. The goal of this work is to share with the reader the conceptual basis and a careful theoretical analysis of the q-space technique from first principles and show the fundamental relationship between the q-space technique, the diffusion tensor and the Fick's laws.

We also presented a careful analysis of the q-space technique in the large dynamic displacement limit and showed its connection to the diffusion tensor. Interested readers are encouraged to consult other excellent pedagogical accounts of diffusion NMR and MRI in this journal, see examples Refs([14-18]).

In the process of establishing the fundamental relationship between the diffusion tensor and the q-space technique, we found the correct relationship between the negative Hessian matrix of the real part of the normalized three-dimensional q-space signal and the root-mean-squared displacement tensor derived from three-dimensional anisotropic diffusion tensor equation.

Recent developments in the q-space technique are extensive and very exciting. Great advances have been made in the construction of novel measures[19-20] from the average propagator as well as the development of novel representations[20-22] of the average propagator. Constrained estimation[20] of the average propagator played an important role in extracting accurate average propagator from noisy data. Optimal acquisition[23-26] of three-dimensional q-space measurements is another research of great interest.

**Appendix A: Solution to the one-dimensional diffusion equation**

In this appendix, we will derive the solution to the one-dimensional diffusion equation of the following form:

$$\frac{\partial P(x,t)}{\partial t} = D \frac{\partial^2 P(x,t)}{\partial x^2},$$ (A1)

or

$$\frac{\partial P(x,t)}{\partial t} - D \frac{\partial^2 P(x,t)}{\partial x^2} = 0.$$ (A2)

As mentioned in Section 1, we can think of the probability density function $P(x,t)$ and the molecular concentration $C(x,t)$ interchangeably since $P(x,t) = C(x,t)/M$. Hence, we will use the probability density function $P(x,t)$ throughout the appendices. The approach to be presented here is known as the Fourier method, which also involves the method of separation of variables. Before we begin with the derivation, a remark on notation is in order. The convention of our notation will be used throughout the appendices. For convenience, the Fourier transform pair of the probability density function with respect to the spatial coordinate systems, $x$ and $k$, is given by:

$$P(x,t) = \int_{-\infty}^{+\infty} \widetilde{P}(k,t) e^{+i 2\pi k x} dk,$$ (A3)

and

$$\widetilde{P}(k,t) = \int_{-\infty}^{+\infty} P(x,t) e^{-i 2\pi k x} dx$$ (A4)

It is easy to see that the second order spatial derivative and the first order temporal derivative of $P(x,t)$ are given by:

$$\frac{\partial^2 P(x,t)}{\partial x^2} = \int_{-\infty}^{+\infty} \widetilde{P}(k,t) \left( -(2\pi k)^2 \right) e^{+i 2\pi k x} dk,$$ (A5)

and

$$\frac{\partial P(x,t)}{\partial t} = \int_{-\infty}^{+\infty} \frac{\partial \widetilde{P}(k,t)}{\partial t} e^{+i2\pi kx} dk \qquad\qquad (A6)$$

Substituting Eq. [A5] and Eq. [A6] into Eq. [A2] leads to a first-order partial differential equation:

$$\frac{\partial P(x,t)}{\partial t} - D\frac{\partial^2 P(x,t)}{\partial x^2} = \int_{-\infty}^{+\infty} \left[ \frac{\partial \widetilde{P}(k,t)}{\partial t} + D(2\pi k)^2 \widetilde{P}(k,t) \right] e^{+i2\pi kx} dk = 0 \qquad (A7)$$

or

$$\frac{\partial \widetilde{P}(k,t)}{\partial t} + D(2\pi k)^2 \widetilde{P}(k,t) = 0 \qquad\qquad (A8)$$

By the method of separation of variables, we shall let $\widetilde{P}(k,t)$ be given by:

$$\widetilde{P}(k,t) = K(k)T(t). \qquad\qquad (A9)$$

Substituting Eq. [A9] into Eq. [A8] leads to

$$K(k)\left( \frac{dT(t)}{dt} + D(2\pi k)^2 T(t) \right) = 0. \qquad\qquad (A10)$$

Because $K(k)$ is an arbitrary function and is proportional to $\widetilde{P}(k,t)$, it cannot be a null function.

Therefore, the non-trivial solution to Eq. [A10] is the following ordinary differential equation:

$$\frac{dT(t)}{dt} + D(2\pi k)^2 T(t) = 0, \qquad\qquad (A11)$$

and its solution is given by:

$$T(t) = Ae^{-D(2\pi k)^2 t}. \qquad\qquad (A12)$$

Since $\widetilde{P}(k,t=0) = K(k)A$ at $t=0$, $\widetilde{P}(k,t)$ can be expressed conveniently as

$$\widetilde{P}(k,t) = \widetilde{P}(k,0)e^{-D(2\pi k)^2 t}. \qquad\qquad (A13)$$

By substituting Eq. [A13] in Eq. [A3], we arrive at

$$P(x,t) = \int_{-\infty}^{+\infty} \underbrace{\widetilde{P}(k,0)}_{\int_{-\infty}^{+\infty} P(x',0) e^{-i2\pi k x'} dx'} e^{-D(2\pi k)^2 t} e^{+i2\pi k x} dk$$

$$= \int_{-\infty}^{+\infty} P(x',0) \underbrace{\left( \int_{-\infty}^{+\infty} e^{-D(2\pi k)^2 t} e^{+i2\pi k (x-x')} dk \right)}_{\dfrac{1}{(4\pi Dt)^{1/2}} e^{-\dfrac{(x-x')^2}{4Dt}}} dx'$$

$$= \frac{1}{(4\pi Dt)^{1/2}} \int_{-\infty}^{+\infty} P(x',0) \, e^{-\frac{(x-x')^2}{4Dt}} dx'. \tag{A14}$$

Note $P(x',0)$ is the initial condition of the probability density function at $x'$ and at $t=0$ or $C(x',0)$ is the molecular concentration at $x'$ and at $t=0$. In the context of diffusion MRI, $t=0$ is referred to the time when the first gradient pulse is applied. As explained in Section 4, it is conceivable that the projection of the initial probability (or concentration) of the tagged diffusible water molecules, $P(x',0)$ (or $C(x',0)$), along the axis of the diffusion gradient direction closely resembles that of the Dirac delta distribution. Suppose the initial concentration is modeled as the Dirac delta distribution, i.e., $P(x',0) = \delta(x')$ (or $C(x',0) = M\delta(x')$), then Eq. [A14] reduces to

$$P(x,t) = \frac{1}{(4\pi Dt)^{1/2}} \exp(-\frac{x^2}{4Dt}), \tag{A15}$$

or

$$C(x,t) = \frac{M}{(4\pi Dt)^{1/2}} \exp(-\frac{x^2}{4Dt}).$$

$P(x,t)$ is the Gaussian probability density function of the displacement $x$ with the standard deviation given by the root-mean-squared displacement, $x_{rms}$, or the second moment of the Gaussian probability density function:

$$x_{rms} \equiv \sqrt{\left\langle x^2 \right\rangle} = \sqrt{\int_{-\infty}^{\infty} x^2 P(x,t) dx} \tag{A16}$$

$$= \sqrt{\int_{-\infty}^{\infty} x^2 \frac{1}{(4\pi Dt)^{1/2}} \exp(-\frac{x^2}{4Dt}) dx}$$

$$= \sqrt{2Dt} \,. \tag{A17}$$

**Appendix B: Solution to the three-dimensional isotropic diffusion equation**

The solution to the three-dimensional isotropic diffusion equation is very similar to the one-dimensional case. We shall provide only the essential steps in the derivation. The three-dimensional isotropic diffusion equation is given by:

$$\frac{\partial P(\mathbf{r},t)}{\partial t} = D\nabla^2 P(\mathbf{r},t) = D\left(\frac{\partial^2 P(\mathbf{r},t)}{\partial x^2} + \frac{\partial^2 P(\mathbf{r},t)}{\partial y^2} + \frac{\partial^2 P(\mathbf{r},t)}{\partial z^2}\right),$$  (B1)

with $\mathbf{r} = [x, y, z]^T$. Let $\mathbf{k} = [k_x, k_y, k_z]^T$, the Fourier transform pair of $C(\mathbf{r},t)$ is given by:

$$P(\mathbf{r},t) = \iiint \widetilde{P}(\mathbf{k},t)\exp(+i2\pi\mathbf{k}\cdot\mathbf{r})\,dk_x\,dk_y\,dk_z,$$  (B2)

and

$$\widetilde{P}(\mathbf{k},t) = \iiint P(\mathbf{r},t)\exp(-i2\pi\mathbf{k}\cdot\mathbf{r})\,dx\,dy\,dz.$$  (B3)

Note that the limits of integration are on the entire real axis and this notation will be used throughout this work.

Upon taking the second order spatial derivatives and the first order temporal derivative of $P(\mathbf{r},t)$ in Eq. [B2] and Eq. [B3], we are led to the first-order partial differential equation of the following form:

$$\frac{\partial \widetilde{P}(\mathbf{k},t)}{\partial t} + (2\pi)^2(k_x^2 + k_y^2 + k_z^2)D\widetilde{P}(\mathbf{k},t) = 0$$  (B4)

By the method of separation of variables with the following expression for $\widetilde{P}(\mathbf{k},t)$:

$$\widetilde{P}(\mathbf{k},t) = K(\mathbf{k})T(t),$$

and by substituting this expression in Eq. [B4], we are led to the first-order ordinary differential equation in time whose solution is given by:

$$T(t) = A\exp(-(2\pi)^2(k_x^2 + k_y^2 + k_z^2)Dt)$$  (B5)

With the use of initial condition of $\widetilde{P}(\mathbf{k},0) = K(\mathbf{k})T(0) = K(\mathbf{k})A$, $\widetilde{P}(\mathbf{k},t)$ can be expressed as:

$$\widetilde{P}(\mathbf{k},t) = \widetilde{P}(\mathbf{k},0)\exp(-(2\pi)^2(k_x^2 + k_y^2 + k_z^2)Dt).$$ (B6)

Substituting Eq. [B6] into Eq. [B2], we arrive at

$$P(\mathbf{r},t) = \iiint \left( \iiint \exp(-(2\pi)^2(\mathbf{k}^T \cdot \mathbf{k})Dt)\exp(+i2\pi\mathbf{k}\cdot(\mathbf{r}-\mathbf{r}'))dk_x dk_y dk_z \right) P(\mathbf{r}',0)dx'dy'dz'$$ (B7)

and

$$P(\mathbf{r},t) = \frac{1}{(4\pi Dt)^{3/2}} \iiint P(\mathbf{r}',0)\exp(-\frac{(x-x')^2 + (y-y')^2 + (z-z')^2}{4Dt})dx'dy'dz'.$$

(B8)

If we assume that $P(\mathbf{r}',0) = \delta(\mathbf{r}')$, Eq. [B8] reduces to

$$P(\mathbf{r},t) = \frac{1}{(4\pi Dt)^{3/2}}\exp(-\frac{x^2+y^2+z^2}{4Dt}).$$ (B9)

The root-mean-square displacement, $r_{rms}$, is then given by:

$$r_{rms} \equiv \sqrt{\langle \mathbf{r}^T \cdot \mathbf{r} \rangle} = \sqrt{\iiint (x^2 + y^2 + z^2)P(\mathbf{r},t)dxdydz}$$

$$= \sqrt{\iiint (x^2 + y^2 + z^2)\frac{1}{(4\pi Dt)^{3/2}}\exp(-\frac{x^2+y^2+z^2}{4Dt})\,dxdydz}$$

$$= \sqrt{6Dt}.$$ (B10)

**Appendix C: Solution to the three-dimensional anisotropic diffusion equation**

The three-dimensional anisotropic diffusion equation given in Eq. [8] can be expressed in matrix form as:

$$\frac{\partial P(\mathbf{r},t)}{\partial t} = \begin{pmatrix} \frac{\partial}{\partial x} & \frac{\partial}{\partial y} & \frac{\partial}{\partial z} \end{pmatrix} \begin{pmatrix} D_{xx} & D_{xy} & D_{xz} \\ D_{xy} & D_{yy} & D_{yz} \\ D_{xz} & D_{yz} & D_{zz} \end{pmatrix} \begin{pmatrix} \frac{\partial}{\partial x} \\ \frac{\partial}{\partial y} \\ \frac{\partial}{\partial z} \end{pmatrix} P(\mathbf{r},t) \,. \tag{C1}$$

Since the diffusion tensor is assumed to be symmetric, it can be diagonalized by orthogonal transformation through the eigenvalue decomposition:

$$\mathbf{D} = \mathbf{R}\boldsymbol{\Lambda}\mathbf{R}^T \,,$$

where the orthogonal matrix $\mathbf{R}$ is a proper rotation matrix, i.e., $\mathbf{R}\mathbf{R}^T = \mathbf{I}$ and $\det(\mathbf{R}) = 1$, and the eigenvectors of $\mathbf{D}$ are arranged as column vectors of $\mathbf{R}$. By a change of variables from $\mathbf{r} = [x, y, z]^T$ to $\boldsymbol{\eta} = [\eta_x, \eta_y, \eta_z]^T$ by the following form:

$$\nabla_{\boldsymbol{\eta}} \equiv \begin{pmatrix} \frac{\partial}{\partial \eta_x} \\ \frac{\partial}{\partial \eta_y} \\ \frac{\partial}{\partial \eta_z} \end{pmatrix} = \underbrace{\begin{pmatrix} \frac{\partial x}{\partial \eta_x} & \frac{\partial y}{\partial \eta_x} & \frac{\partial z}{\partial \eta_x} \\ \frac{\partial x}{\partial \eta_y} & \frac{\partial y}{\partial \eta_y} & \frac{\partial z}{\partial \eta_y} \\ \frac{\partial x}{\partial \eta_z} & \frac{\partial y}{\partial \eta_z} & \frac{\partial z}{\partial \eta_z} \end{pmatrix}}_{\mathbf{G}_{\boldsymbol{\eta}}(\mathbf{r})} \begin{pmatrix} \frac{\partial}{\partial x} \\ \frac{\partial}{\partial y} \\ \frac{\partial}{\partial z} \end{pmatrix} \equiv \boldsymbol{\Lambda}^{1/2}\mathbf{R}^T\nabla_{\mathbf{r}} \,, \tag{C2}$$

or

$$d\boldsymbol{\eta} \equiv \begin{pmatrix} d\eta_x \\ d\eta_y \\ d\eta_z \end{pmatrix} = \underbrace{\begin{pmatrix} \frac{\partial \eta_x}{\partial x} & \frac{\partial \eta_x}{\partial y} & \frac{\partial \eta_x}{\partial z} \\ \frac{\partial \eta_y}{\partial x} & \frac{\partial \eta_y}{\partial y} & \frac{\partial \eta_y}{\partial z} \\ \frac{\partial \eta_z}{\partial x} & \frac{\partial \eta_z}{\partial y} & \frac{\partial \eta_z}{\partial z} \end{pmatrix}}_{\mathbf{J}_{\mathbf{r}}(\boldsymbol{\eta})} \begin{pmatrix} dx \\ dy \\ dz \end{pmatrix} = \boldsymbol{\Lambda}^{-1/2}\mathbf{R}^T d\mathbf{r} \,. \tag{C3}$$

Note that

$$\nabla_{\boldsymbol{\eta}} \equiv \begin{pmatrix} \dfrac{\partial}{\partial \eta_x} \\[2mm] \dfrac{\partial}{\partial \eta_y} \\[2mm] \dfrac{\partial}{\partial \eta_z} \end{pmatrix} = \begin{pmatrix} \dfrac{\partial x}{\partial \eta_x} & \dfrac{\partial y}{\partial \eta_x} & \dfrac{\partial z}{\partial \eta_x} \\[2mm] \dfrac{\partial x}{\partial \eta_y} & \dfrac{\partial y}{\partial \eta_y} & \dfrac{\partial z}{\partial \eta_y} \\[2mm] \dfrac{\partial x}{\partial \eta_z} & \dfrac{\partial y}{\partial \eta_z} & \dfrac{\partial z}{\partial \eta_z} \end{pmatrix} \begin{pmatrix} \dfrac{\partial}{\partial x} \\[2mm] \dfrac{\partial}{\partial y} \\[2mm] \dfrac{\partial}{\partial z} \end{pmatrix} = \underbrace{\begin{pmatrix} \dfrac{\partial x}{\partial \eta_x} & \dfrac{\partial y}{\partial \eta_x} & \dfrac{\partial z}{\partial \eta_x} \\[2mm] \dfrac{\partial x}{\partial \eta_y} & \dfrac{\partial y}{\partial \eta_y} & \dfrac{\partial z}{\partial \eta_y} \\[2mm] \dfrac{\partial x}{\partial \eta_z} & \dfrac{\partial y}{\partial \eta_z} & \dfrac{\partial z}{\partial \eta_z} \end{pmatrix} \begin{pmatrix} \dfrac{\partial \eta_x}{\partial x} & \dfrac{\partial \eta_y}{\partial x} & \dfrac{\partial \eta_z}{\partial x} \\[2mm] \dfrac{\partial \eta_x}{\partial y} & \dfrac{\partial \eta_y}{\partial y} & \dfrac{\partial \eta_z}{\partial y} \\[2mm] \dfrac{\partial \eta_x}{\partial z} & \dfrac{\partial \eta_y}{\partial z} & \dfrac{\partial \eta_z}{\partial z} \end{pmatrix}}_{\mathbf{I}} \begin{pmatrix} \dfrac{\partial}{\partial \eta_x} \\[2mm] \dfrac{\partial}{\partial \eta_y} \\[2mm] \dfrac{\partial}{\partial \eta_z} \end{pmatrix},$$

and

$$d\boldsymbol{\eta} \equiv \begin{pmatrix} d\eta_x \\ d\eta_y \\ d\eta_z \end{pmatrix} = \begin{pmatrix} \dfrac{\partial \eta_x}{\partial x} & \dfrac{\partial \eta_x}{\partial y} & \dfrac{\partial \eta_x}{\partial z} \\[2mm] \dfrac{\partial \eta_y}{\partial x} & \dfrac{\partial \eta_y}{\partial y} & \dfrac{\partial \eta_y}{\partial z} \\[2mm] \dfrac{\partial \eta_z}{\partial x} & \dfrac{\partial \eta_z}{\partial y} & \dfrac{\partial \eta_z}{\partial z} \end{pmatrix} \begin{pmatrix} dx \\ dy \\ dz \end{pmatrix} = \underbrace{\begin{pmatrix} \dfrac{\partial \eta_x}{\partial x} & \dfrac{\partial \eta_x}{\partial y} & \dfrac{\partial \eta_x}{\partial z} \\[2mm] \dfrac{\partial \eta_y}{\partial x} & \dfrac{\partial \eta_y}{\partial y} & \dfrac{\partial \eta_y}{\partial z} \\[2mm] \dfrac{\partial \eta_z}{\partial x} & \dfrac{\partial \eta_z}{\partial y} & \dfrac{\partial \eta_z}{\partial z} \end{pmatrix} \begin{pmatrix} \dfrac{\partial x}{\partial \eta_x} & \dfrac{\partial x}{\partial \eta_y} & \dfrac{\partial x}{\partial \eta_z} \\[2mm] \dfrac{\partial y}{\partial \eta_x} & \dfrac{\partial y}{\partial \eta_y} & \dfrac{\partial y}{\partial \eta_z} \\[2mm] \dfrac{\partial z}{\partial \eta_x} & \dfrac{\partial z}{\partial \eta_y} & \dfrac{\partial z}{\partial \eta_z} \end{pmatrix}}_{\mathbf{I}} \begin{pmatrix} d\eta_x \\ d\eta_y \\ d\eta_z \end{pmatrix},$$

therefore, we have the following properties

1. $\mathbf{G}_{\boldsymbol{\eta}}(\mathbf{r})\mathbf{G}_{\mathbf{r}}(\boldsymbol{\eta}) = \mathbf{I}$,

2. $\mathbf{J}_{\mathbf{r}}(\boldsymbol{\eta})\mathbf{J}_{\boldsymbol{\eta}}(\mathbf{r}) = \mathbf{I}$,

3. $\mathbf{G}_{\mathbf{r}}(\boldsymbol{\eta}) = \mathbf{J}_{\mathbf{r}}^{T}(\boldsymbol{\eta})$,

4. $\mathbf{G}_{\boldsymbol{\eta}}(\mathbf{r}) = \mathbf{J}_{\boldsymbol{\eta}}^{T}(\mathbf{r})$,

5. $\mathbf{G}_{\boldsymbol{\eta}}(\mathbf{r}) = \mathbf{G}_{\mathbf{r}}^{-1}(\boldsymbol{\eta})$, and

6. $\mathbf{J}_{\mathbf{r}}(\boldsymbol{\eta}) = \mathbf{J}_{\boldsymbol{\eta}}^{-1}(\mathbf{r})$.

Note that the right-hand side of Eq. [C1] is derived from the above properties and $\mathbf{J}_{\mathbf{r}}(\boldsymbol{\eta})$ is the Jacobian matrix of the $\boldsymbol{\eta}$-coordinate system with respect to the $\mathbf{r}$-coordinate system. The coordinate transformation used here is very similar to that used in our recent work on the analytical error propagation framework for diffusion tensor imaging[27-29].

We can change the diffusion equation in the $\mathbf{r}$-coordinate system

$$\frac{\partial P(\mathbf{r},t)}{\partial t} = \nabla_{\mathbf{r}} \cdot \left( \mathbf{D} \nabla_{\mathbf{r}} P(\mathbf{r},t) \right)$$

into

$$\frac{\partial \hat{P}(\boldsymbol{\eta},t)}{\partial t} = \nabla_{\boldsymbol{\eta}}^2 \hat{P}(\boldsymbol{\eta},t) \, , \tag{C4}$$

which is the diffusion equation in the $\boldsymbol{\eta}$-coordinate system. Note that $\nabla_{\mathbf{r}}$ is the gradient operator with respect to the $\mathbf{r}$-coordinate system. Similarly, $\nabla_{\boldsymbol{\eta}}^2$ is the Laplacian operator with respect to the $\boldsymbol{\eta}$-coordinate system. Further, it is important to point out that the correspondence between $\hat{P}(\boldsymbol{\eta},t)$ and $P(\mathbf{r},t)$, which is given by

$$\hat{P}(\boldsymbol{\eta},t) = \hat{P}(\Lambda^{-1/2} \mathbf{R}^T \mathbf{r},t) = P(\mathbf{r},t)$$

The validity of $\boldsymbol{\eta} = \Lambda^{-1/2} \mathbf{R}^T \mathbf{r}$ is based upon constancy of the diffusion tensor and it is derived from Eq. [C3]. It is interesting to note that diffusion is isotropic in the $\boldsymbol{\eta}$-coordinate system and the diffusion coefficient has the value of unity. The solution to Eq. [C4] can be gleaned from Appendix B by setting $D$ to unity; that is,

$$\hat{P}(\boldsymbol{\eta},t) = \frac{1}{(4\pi t)^{3/2}} \iiint \hat{P}(\boldsymbol{\eta}',0) \exp\left(-\frac{(\eta_x - \eta_x')^2 + (\eta_y - \eta_y')^2 + (\eta_z - \eta_z')^2}{4t}\right) d\eta_x' d\eta_y' d\eta_z'$$

or

$$\hat{P}(\boldsymbol{\eta},t) = \frac{1}{(4\pi t)^{3/2}} \iiint \hat{P}(\boldsymbol{\eta}',0) \exp\left(-\frac{\|\boldsymbol{\eta} - \boldsymbol{\eta}'\|^2}{4t}\right) d\eta_x' d\eta_y' d\eta_z' \, .$$

We are now ready to transform the integral above to the $\mathbf{r}$-coordinate system with the aid of the determinant of the Jacobian matrix as follows:

$$d\eta_x d\eta_y d\eta_z = \left| \det(\mathbf{J}_{\mathbf{r}}(\boldsymbol{\eta})) \right| dx dy dz \, ,$$

$$= \left| \det(\mathbf{\Lambda}^{-1/2} \mathbf{R}^T) \right| dx\,dy\,dz$$

$$= \left| \det(\mathbf{\Lambda}^{-1/2}) \right| \left| \det(\mathbf{R}^T) \right| dx\,dy\,dz$$

$$= \frac{1}{\sqrt{\lambda_1 \lambda_2 \lambda_3}} dx\,dy\,dz \,,$$

where $\lambda_1$, $\lambda_2$, and $\lambda_3$ are the eigenvalues of the diffusion tensor. The resultant integral in the $\mathbf{r}$-coordinate system is given by:

$$\hat{P}(\mathbf{\eta}, t) = \frac{1}{(4\pi t)^{3/2}} \iiint \hat{P}(\mathbf{\eta'}, 0) \exp\left(-\frac{\|\mathbf{\eta} - \mathbf{\eta'}\|^2}{4t}\right) d\eta_x' d\eta_y' d\eta_z' \,,$$

$$\hat{P}(\mathbf{\Lambda}^{-1/2}\mathbf{R}^T\mathbf{r}, t) = \frac{1}{(4\pi t)^{3/2}} \frac{1}{\sqrt{\lambda_1 \lambda_2 \lambda_3}} \iiint \hat{P}(\mathbf{\Lambda}^{-1/2}\mathbf{R}^T\mathbf{r'}, 0) \exp\left(-\frac{\left\|\mathbf{\Lambda}^{-1/2}\mathbf{R}^T(\mathbf{r} - \mathbf{r'})\right\|^2}{4t}\right) dx' dy' dz'$$

$$C(\mathbf{r}, t) = \hat{P}(\mathbf{\Lambda}^{-1/2}\mathbf{R}^T\mathbf{r}, t) \,,$$

$$= \frac{1}{(4\pi t)^{3/2}} \frac{1}{\sqrt{\det(\mathbf{D})}} \iiint P(\mathbf{r'}, 0) \exp\left(-\frac{(\mathbf{r} - \mathbf{r'})^T \mathbf{R} \mathbf{\Lambda}^{-1} \mathbf{R}^T (\mathbf{r} - \mathbf{r'})}{4t}\right) dx' dy' dz' \,,$$

$$= \frac{1}{(4\pi t)^{3/2}} \frac{1}{\sqrt{\det(\mathbf{D})}} \iiint P(\mathbf{r'}, 0) \exp\left(-\frac{(\mathbf{r} - \mathbf{r'})^T \mathbf{D}^{-1} (\mathbf{r} - \mathbf{r'})}{4t}\right) dx' dy' dz' \,. \tag{C5}$$

If we again assume that $P(\mathbf{r'}, 0) = \delta(\mathbf{r'})$, then Eq. [C5] reduces to:

$$P(\mathbf{r}, t) = \frac{1}{(4\pi t)^{3/2}} \frac{1}{\sqrt{\det(\mathbf{D})}} \exp\left(-\frac{\mathbf{r}^T \mathbf{D}^{-1} \mathbf{r}}{4t}\right) \,. \tag{C6}$$

In the case of anisotropic diffusion, the mean-squared displacement tensor is a matrix of the following form:

$$\left\langle \mathbf{r}\mathbf{r}^T \right\rangle = \iiint \begin{pmatrix} x^2 & xy & xz \\ xy & y^2 & yz \\ xz & yz & z^2 \end{pmatrix} P(\mathbf{r},t)\,dx\,dy\,dz \ . \tag{C7}$$

Note that the integration is component-wise. Again, if we adopt the change of variables similar to the above, i.e., $\boldsymbol{\eta} = \boldsymbol{\Lambda}^{-1/2}\mathbf{R}^T\mathbf{r}$,

$$\left\langle \mathbf{r}\mathbf{r}^T \right\rangle = \frac{1}{(4\pi t)^{3/2}} \mathbf{R}\,\boldsymbol{\Lambda}^{1/2} \left[ \iiint \begin{pmatrix} \eta_x^2 & \eta_x\eta_y & \eta_x\eta_z \\ \eta_x\eta_y & \eta_y^2 & \eta_y\eta_z \\ \eta_x\eta_z & \eta_y\eta_z & \eta_z^2 \end{pmatrix} \exp(-\frac{\|\boldsymbol{\eta}\|^2}{4t})\,d\eta_x\,d\eta_y\,d\eta_z \right] \boldsymbol{\Lambda}^{1/2}\,\mathbf{R}^T$$

$$= \frac{1}{(4\pi t)^{3/2}} \mathbf{R}\,\boldsymbol{\Lambda}^{1/2} \begin{pmatrix} 2^4\pi^{3/2}t^{5/2} & 0 & 0 \\ 0 & 2^4\pi^{3/2}t^{5/2} & 0 \\ 0 & 0 & 2^4\pi^{3/2}t^{5/2} \end{pmatrix} \boldsymbol{\Lambda}^{1/2}\,\mathbf{R}^T$$

$$= \frac{2^4\pi^{3/2}t^{5/2}}{(4\pi t)^{3/2}} \mathbf{R}\,\boldsymbol{\Lambda}\,\mathbf{R}^T$$

$$= 2\mathbf{D}t \ . \tag{C8}$$

Finally, we have

$$tr\left(\left\langle \mathbf{r}\mathbf{r}^T \right\rangle\right) = 2\,tr(\mathbf{D})t, \tag{C9}$$

where $tr(\mathbf{D})$ is the trace of $\mathbf{D}$. Note that, Eq. [C9] coincides with Eq. [B10] when $\mathbf{D}$ is isotropic.

## Appendix D: Stejskal-Tanner pulse sequence and its two effective gradient pulse sequences

In Section 3, we established through first principles that the normalized q-space signal, $E(\mathbf{q}, \Delta)$, is related to the average propagator, $P_D(\mathbf{R}, \Delta)$,

$$E(\mathbf{q}, \Delta) = \int P_D(\mathbf{R}, \Delta) \, \exp(-i\, 2\pi \mathbf{q} \cdot \mathbf{R}) d\mathbf{R} \,. \tag{D1}$$

In the process of establishing this important relationship, we mentioned that the effect of the refocusing RF pulse in the Stejskal-Tanner pulse sequence, see Figure 2A, on the phase of the magnetization vector is to change the sign of its phase. Since the phase or frequency information is determined by

$$\omega(\mathbf{r}(\tau), \tau) = \gamma \mathbf{G}(\tau) \cdot \mathbf{r}(\tau) \,, \tag{D2}$$

it is natural to think of the sign change in the phase as a reversal of the polarity of the first pulse field gradient, see Figure 2B. Therefore, the effective gradient shown in Figure 2B is equivalent to the Stejskal-Tanner pulse sequence. It is interesting to note that the pulse sequence shown in Figure 2C is equivalent to the Stejskal-Tanner pulse sequence when there is no motion, i.e., $\mathbf{r}_1 = \mathbf{r}_2$ in Section 3, in the system. When there is motion in the system, the pulse sequence in Figure 2C is no longer consistent with the Stejskal-Tanner pulse sequence. Based on similar derivation as shown in Section 3, it is easy to show that it leads to the normalized q-space signal of the following form:

$$E(\mathbf{q}, \Delta) = \int P_D(\mathbf{R}, \Delta) \, \exp(+i\, 2\pi \mathbf{q} \cdot \mathbf{R}) d\mathbf{R} \,. \tag{D3}$$

The normalized q-space signal is now the inverse Fourier transform of the average propagator. We should note that the phase in the Fourier kernel is not merely a convention. The difference is fundamental. As discussed in Ref.[30] the difference is inconsequential if the average propagator is symmetric, i.e., when $P_D(\mathbf{R}, \Delta) = P_D(-\mathbf{R}, \Delta)$, but important difference manifests itself when the average propagator is not symmetric, e.g., for diffusion taking place near a boundary. Readers interested in the origins and consequences of the phase in the above Fourier relations are urged to study Refs.([30-31]) for further details.

**Acknowledgment**

CGK dedicates this work to Geik Kee Tan. The authors would like to thank Dr. Peter J. Basser for reading the manuscript and for insightful discussion. CGK was supported in part by NIH IRCMH090912-01 and EÖ was supported by NIH R01MH074794.

FIGURES AND CAPTIONS

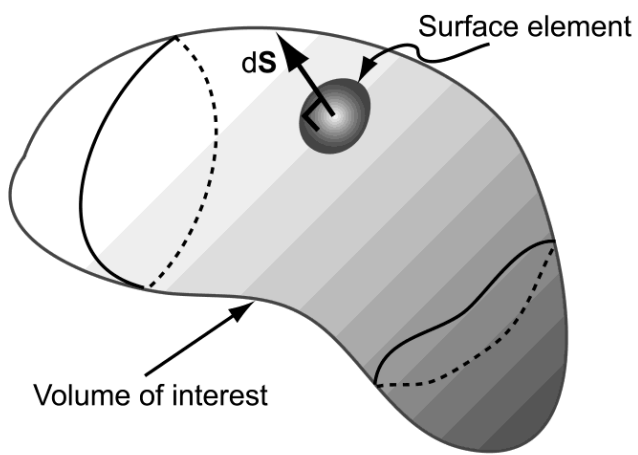

Figure 1. A volume element and an oriented surface element of diffusible molecules.

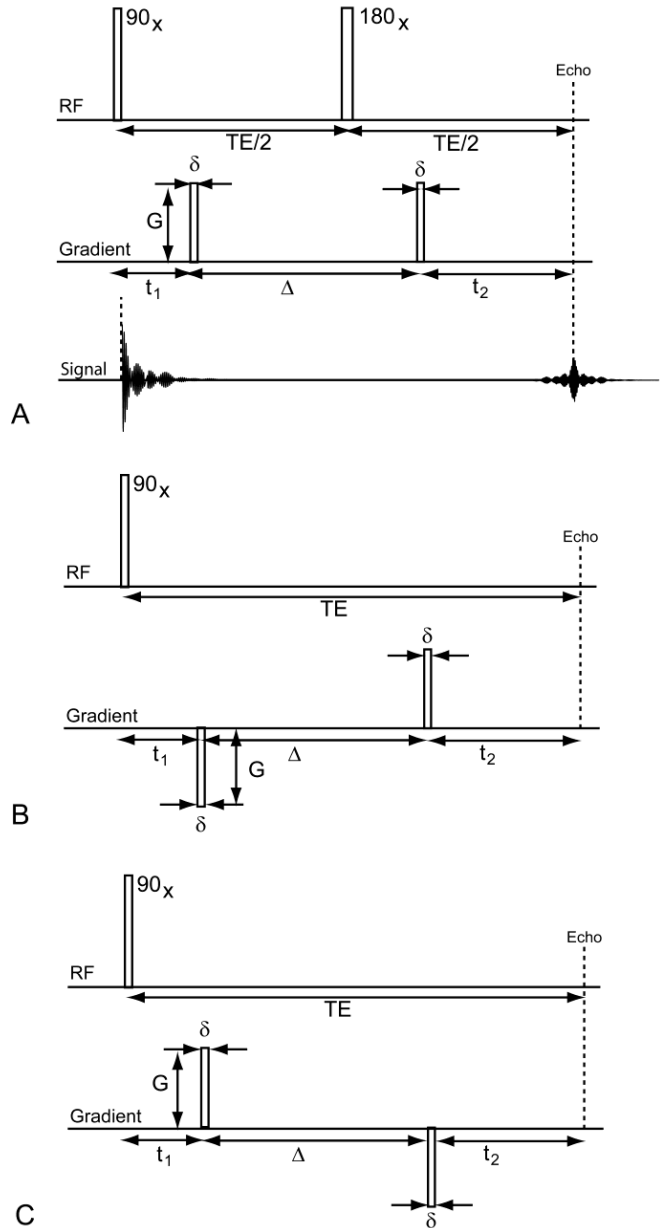

Figure 2. The Stejskal-Tanner pulse sequence (A) and its two effective gradient pulse sequences (B) and (C). In (A), the top diagram is the spin echo pulse sequence with the initial 90 degrees radiofrequency pulse along the x-axis, and after some time TE/2, the 180 degrees radiofrequency pulse along the x-axis. In the middle diagram of panel (A) the application of gradient pulses are shown. The effective gradient pulse sequences, (B) and (C), are equivalent to (A) when there is no motion (i.e., $\mathbf{r}_1 = \mathbf{r}_2$ in Section 3). When translational motion exists, (B) remains consistent with (A) but (C) has a sign difference. The sign difference in (B) resulted in the normalized q-space signal as the inverse Fourier transform of the average propagator while (C), which is inconsistent with (A), resulted in the forward Fourier transform.

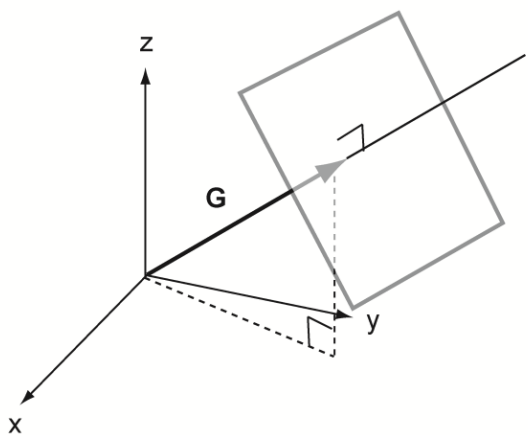

Figure 3. Diffusion gradient pulse in the direction of **G** imparts the same phase, say $\omega_0$, for all water molecules situated (with position vector **r**) on a particular plane perpendicular to **G**. Hence, the concentration of water molecules with the same phase, $\omega_0$, can be thought of as a one dimensional function resembling that of the Dirac delta function along the direction of **G**.